\documentclass[aps,onecolumn,11pt,floatfix,altaffilletter,tightenlines,showpacs,showkeys,notitlepage,nofootinbib]{revtex4-1}
\pdfoutput=1

\usepackage[colorlinks=true,citecolor=blue,linkcolor=blue]{hyperref}
\usepackage[normalem]{ulem}
\usepackage{amsmath,amssymb}
\usepackage{slashed}
\usepackage{nicefrac}
\usepackage{mathtools}
\usepackage{epsfig}
\usepackage{graphicx}             
\usepackage{url}
\usepackage{color}
\usepackage{multirow}
\usepackage{placeins}
\usepackage[dvipsnames]{xcolor}
\usepackage{epstopdf}
\usepackage[abs]{overpic}
\usepackage{csquotes}
\usepackage{comment}

\usepackage[caption=false]{subfig}
\usepackage{siunitx}
\usepackage{pifont}

\usepackage{tikz}
\usetikzlibrary{trees}
\usetikzlibrary{decorations.pathmorphing}
\usetikzlibrary{decorations.markings}

\definecolor{jblue}  {RGB}{20,50,100}
\definecolor{npurple}  {RGB} {153, 51, 204}
\definecolor{wred}   {RGB}{217,0,56}
\definecolor{white}   {RGB}{255,255,255}

\definecolor{korange}   {RGB}{235, 80,  43}
\definecolor{korange2}   {RGB}{245, 100,  63}
\definecolor{kyelloworange}   {RGB}{255, 210,  110}
\definecolor{kyelloworange2}   {RGB}{240, 170,  90}
\definecolor{kred}   {RGB}{204,  102, 153}
\definecolor{kpurple}   {RGB}{153,  61, 190}
\definecolor{kpurplelight}   {RGB}{213,  161, 230}

\usepackage[capitalize]{cleveref}
\usepackage{lipsum}
\definecolor{red}{rgb}{1.0, 0, 0}
 \usepackage{gensymb}
 
\allowdisplaybreaks

\bibpunct{[}{]}{,}{n}{}{,}

\DeclarePairedDelimiter\chevron{\langle}{\rangle}

\newcommand{\I}{\mathrm{i}}
\newcommand{\E}{\mathrm{e}}
\newcommand{\ie}{i.e.~}
\newcommand{\eg}{e.g.~}
\newcommand{\cf}{cf.~}

\newcommand{\tinytext}[1]{\text{\tiny{#1}}}
\newcommand{\fineq}[1]{\;{#1}}

\newcommand{\MSbar}[0]{\overline{\text{MS}}}

\newcommand{\dd}{\mbox{d}}	% differential d
\newcommand{\GWscale}{\Lambda}
\newcommand{\MPl}{M_\tinytext{Pl}}
\newcommand{\Sthree}{\mathcal{S}_3}
\newcommand{\yM}{y_\tinytext{M}}

\newcommand{\symhspace}[2]{\hspace{#1}#2\hspace{#1}}

\pacs{}
\keywords{}

\begin{document}

%=============================================================================
\title{Gravitational Waves from First-Order Phase Transitions:\\ LIGO as a Window
to Unexplored Seesaw Scales}

\author{Vedran Brdar$^{1}$}   \email{vbrdar@mpi-hd.mpg.de}
\author{Alexander J.~Helmboldt$^{1}$} \email{alexander.helmboldt@mpi-hd.mpg.de}
\author{Jisuke Kubo$^{1,2\,}$} \email{kubo@mpi-hd.mpg.de}
\affiliation{$^1$Max-Planck-Institut f\"ur Kernphysik,
       69117~Heidelberg, Germany\\
$^2$Department of Physics, University of Toyama, 3190 Gofuku, Toyama 930-8555, Japan}

\begin{abstract}
\noindent
Within a recently proposed classically conformal model, in which the
generation of neutrino masses is linked to spontaneous scale symmetry
breaking, we investigate the associated phase transition and find it to
be of strong first order with a substantial amount of supercooling.
Carefully taking into account the vacuum energy of the meta-stable minimum, we demonstrate that a significant fraction of the model's parameter space can be excluded simply because the phase transition cannot complete.
We argue this to be a powerful consistency check applicable to general theories based on classical scale invariance.
Finally, we show that all remaining parameter points predict a sizable gravitational wave signal, so that the model can be fully tested by future gravitational wave observatories.
In particular, most of the parameter space can already be probed by the upcoming LIGO science run starting in early 2019.
\end{abstract}

\maketitle

%-----------------------------------------------------------------------------
\section{Introduction}
\label{sec:intro}
%=======
\noindent
%=======
%
% see-saw mechanism
%
% motivation for right-handed neutrinos
Right-handed neutrinos are among the best-motivated particles of beyond-the-Standard Model physics. Their role in explaining light active neutrinos \cite{Minkowski,Yanagida:1979as,GellMann:1980vs,Goran}, the baryon asymmetry of the universe \cite{Fukugita:1986hr}, or dark matter \cite{Asaka:2005an} makes them a key player in many theories of new physics.
% undetermined mass scale
However, their mass scale remains a priori completely unclear.
In fact, reflecting the aforementioned variety in applications, well-motivated scenarios suggest masses ranging from a few keV  all the way up to the GUT scale at about \SI{e16}{GeV} \cite{Drewes:2013gca}.
% experimental problem for too heavy neutrinos
Obviously, if right-handed neutrinos are above the TeV scale, their \textit{direct} detection at current or near-future colliders is not possible.
Alternative experimental strategies allowing to test such high-mass scenarios are thus highly desirable.

% neutrinos and gravitational waves
One such approach is motivated by the landmark discovery of gravitational waves in 2016 and 2017 \cite{Abbott2016,TheLIGOScientific:2017qsa,GBM:2017lvd}:
Since right-handed neutrino masses may be generated in a first-order cosmic phase transition, they may well be linked to the production of a stochastic gravitational wave background observable in current or future experiments \cite{Okada:2018xdh,Long:2017rdo}.
Thus, a complementary \textit{indirect} handle on probing the aforementioned mass scale becomes available.

%
% role of see-saw mechanism in radiatively generating the Higgs potential
%
In this context, the present article will discuss a recently proposed class of models, which suggests yet another potential virtue of (heavy) right-handed neutrinos.
Namely, the authors of Refs. \cite{Brivio:2017dfq,Brivio:2018rzm} demonstrated that right-handed neutrinos with Majorana masses of order \SI{e8}{GeV} may also play a crucial role in stabilizing the Higgs mass and thus in solving the infamous gauge hierarchy problem.
To be slightly more precise, it was proven that in the absence of a bare Higgs mass term, the Standard Model scalar potential can still be correctly reproduced via radiative corrections due to integrating out the heavy right-handed neutrino degrees of freedom.
The large hierarchy between the electroweak and the Majorana mass scale is thus explained in a technically natural way \cite{tHooft:1979rat} by linking it to the smallness of the Dirac neutrino Yukawa coupling.

However, the underlying concept of neglecting one explicit mass term while retaining another is, of course, not renormalization group invariant, and hence calls for a symmetry-based justification, ideally in a UV-complete framework, which does not reintroduce severe parameter fine tuning.
Along these lines, a minimal renormalizable model based on classical scale invariance has recently been proposed in Ref.~\cite{Brdar:2018vjq}.
Here, the Majorana mass scale is generated spontaneously from a conformal tree-level Lagrangian via perturbative quantum effects.

%
% GW signal
%
In the present paper, we will show that the associated cosmic phase transition \textendash\ hereafter referred to as the scale phase transition \textendash\ is necessarily of strong first order and thus involves a considerable amount of supercooling.
The corresponding gravitational wave signal is therefore expected to be sufficiently strong to be testable at both existing and planned observatories.
Interestingly, consistency of the model in \cite{Brdar:2018vjq} requires its fundamental scale to be above \SI{e7}{GeV}, suggesting that even ground-based experiments like LIGO can be sensitive \cite{Dev2016}.

% general expectations
Let us finally remark that, even though we will focus on the concrete model of Ref.~\cite{Brdar:2018vjq} in the present study, we expect qualitatively similar results for any realization of the basic idea of Ref.~\cite{Brivio:2017dfq} within a classically scale-invariant extension of the Standard Model.
In all such models the Majorana mass scale of order \SI{e7}{GeV} has to be spontaneously generated in a cosmic phase transition, which is generically anticipated to be of strong first-order in theories based on nearly conformal dynamics \cite{Konstandin:2011dr,Prokopec:2018tnq,Marzola:2017jzl,Jinno:2016knw,Hashino:2018wee}.

%
% structure of the paper
%
The paper is structured as follows.
\cref{sec:nu_option} provides a brief summary of the considered model.
Based on that, we will then study the model's scale phase transition in \cref{sec:PT}.
In doing so, we will particularly stress how requiring that the aforementioned transition actually completes, yields a powerful means of narrowing down the viable parameter space.
Subsequently, in \cref{sec:GW} we calculate the stochastic gravitational wave background resulting from the scale phase transition and compare it with the sensitivities of various experiments \cite{Kawamura:2006up,Seoane:2013qna,Aasi:2013wya,Caprini:2015zlo,Sato:2017dkf}.
We finally discuss our results and conclude in \cref{sec:conclusion}.

%-----------------------------------------------------------------------------
\section{The model}
\label{sec:nu_option}
%=======
\noindent
%=======
In this paper we study gravitational wave signatures from the first-order phase transition in the classically scale-invariant model introduced in Ref.~\cite{Brdar:2018vjq}.
There, the authors provide a conformal framework in which both neutrino masses and the Higgs potential arise due to the presence of right-handed neutrinos. Namely, after the spontaneous breaking of scale invariance in which the Majorana masses of right-handed neutrinos are dynamically generated, the active neutrino masses are obtained through the type-I seesaw mechanism \cite{Goran,Minkowski,GellMann:1980vs,Yanagida:1979as}, whereas the Higgs potential stems from quantum effects, \ie loops of heavy right-handed neutrinos. The latter requires suppression of the tree-level mass term $\mu^2 H^\dagger H$ which is naturally realized in scale-invariant models. The authors of Ref.~\cite{Brdar:2018vjq} applied robust numerical methods to  find whether it is simultaneously possible to: $(i)$ satisfy the cosmological bound on the sum of neutrino masses \cite{Ade:2015xua}, $(ii)$ obtain the low-energy parameters of the Higgs potential not to deviate from their Standard Model (SM) values by more than \SI{1}{\%}, and $(iii)$ have no Landau poles below the Planck scale. The positive result was reported \cite{Brdar:2018vjq} and the foremost upshot was a lower bound on the vacuum expectation value (VEV) of the scalar field $S$ that breaks scale invariance
\begin{align}
v_s \gtrsim 10^7 \,\text{GeV}\,.
\end{align}
Hence, besides the electroweak and the Planck scale, in this framework a novel scale, $v_s$, is introduced and it sets the masses of all beyond the Standard Model degrees of freedom. It is worthwhile noting the absence of the hierarchy problem \cite{Bardeen1995b,Meissner:2007xv,Meissner2009a,Davoudiasl:2014pya}.
Classical conformal symmetry protects the separation between $v_s$ and the Planck scale, whereas small lepton portal Yukawa couplings of the SM Higgs are relating the electroweak scale and $v_s$.
In what follows, we briefly introduce the model \cite{Brdar:2018vjq} by putting an emphasis on the one-loop effective potential containing the Coleman-Weinberg term \cite{CW} as well as the leading thermal contributions \cite{Carrington:1991hz}.

Above the conformal symmetry breaking scale the model's Lagrangian is given by
\begin{align}
	\mathcal{L} \supseteq{}& \tfrac{1}{2}\partial_\mu S \partial^\mu S
	+ \tfrac{1}{2}\partial_\mu R \partial^\mu R
	+ \I \bar{N}_R \slashed{\partial} N_R 
	- V_\text{tree}(H,S,R)
	- \left( \tfrac{1}{2} \yM S \bar{N}_R N_R^c + y_\nu \bar{L} \tilde{H} N_R + \text{h.c.} \right)\fineq{.}
	\label{eq:UVlagn}
\end{align}
Here, $H$ and $L$  are the SM Higgs and lepton doublets, $N_R$ denotes right-handed neutrinos, whereas $S$ and $R$ are gauge singlet real scalars. The coupling $\yM$ parametrizes the 
strength of the Yukawa interaction between $S$ and $N_R$, while $y_\nu$ is a lepton portal coupling crucial for generating both active neutrino masses and the Higgs potential. 
The tree-level scalar potential $V_\text{tree}$ from \cref{eq:UVlagn} reads
\begin{align}
	V_\text{tree}(H,S,R) =
	\lambda (H^\dagger H)^2
	+ \lambda_S S^4
	+ \lambda_R R^4 
	+ \lambda_{HS} S^2 (H^\dagger H)
	+ \lambda_{HR} R^2 (H^\dagger H)
	+ \lambda_{SR} S^2 R^2 \fineq{,}
	\label{eq:UVpotential}
\end{align}
where we implicitly assumed $R$ to have an odd charge under some $\mathbb{Z}_2$ symmetry in order to simplify the expression by only including terms with even number of $R$ fields.

Scale invariance must be broken as otherwise the model would not predict massive particles that have been observed in numerous experiments. We assume that, at the scale $\Lambda$, the $S$ field develops a finite VEV, i.e. the following arrangement is achieved
\begin{align}
	\chevron{H}=\chevron{R} =0 \qquad\text{and}\qquad \chevron{S} \equiv v_s\neq 0\,.
	\label{eq:vevs}
\end{align} 
Applying the analytical procedure introduced by Gildener and Weinberg \cite{GW}, it is possible to obtain \cref{eq:vevs} in a rather elegant way, without 
brute-force numerical treatments. The necessary condition that we obtain reads
\begin{align}
	\lambda_S(\Lambda)=0 \fineq{.}
	\label{eq:condition}
\end{align}

The radiative breaking of conformal symmetry induces $\mathcal{O}(\Lambda)$ masses for $N_R$ and $R$ fields.
For convenience and later usage, we define here field-dependent masses instead
\begin{align}
	m_N(S) = \yM S \qquad\text{and}\qquad
	m_R^2(S) = 2\lambda_{SR} S^2 \fineq{,}
\label{eq:field_dependent}
\end{align}
where the Higgs-field-dependent terms are omitted due to rather small couplings of the new particles to SM ($\lambda_{HR},\,y_\nu\ll \yM, \lambda_{SR}\sim \mathcal{O}(1)$\,; for more information  on the magnitude of the couplings in the model see Table I in  Ref.~\cite{Brdar:2018vjq}).

With the potential that develops a flat direction along the $S$ field axis $(\langle H\rangle =\langle R \rangle =0)$ we infer that the only relevant term in the tree-level potential is $\lambda_S S^4$ which, however, vanishes at the scale $\Lambda$ due to the Gildener-Weinberg requirement in \cref{eq:condition}. Hence, the leading contributions to the scalar potential are realized at the quantum level.
The one-loop, daisy-improved finite temperature effective potential reads
\begin{align}
	V_\text{eff}(S,T) = V_\tinytext{CW}(S) + V_\tinytext{FT}(S,T) + V_\text{daisy}(S,T) \fineq{.}
\label{eq:PT:Veff}
\end{align}
Here, $V_\tinytext{CW}$ is the renormalized Coleman-Weinberg potential which in the $\MSbar$ scheme reads
\begin{align}
V_\tinytext{CW}(S) &= \frac{1}{64\pi^2} \left[
m_R^4(S) \left( \log\frac{m_R^2(S)}{\Lambda^2} - \frac{3}{2} \right)
	- 6 m_N^4(S) \left( \log\frac{m_N^2(S)}{\Lambda^2} - \frac{3}{2} \right)
	\right] \nonumber \\ & \equiv A S^4 + B S^4 \log\left(\frac{S^2}{\Lambda^2}\right) \fineq{,}
	\label{eq:PT:VCW}
\end{align}
where the field-dependent tree-level masses from \cref{eq:field_dependent} and the Gildener-Weinberg functions \cite{GW},
\begin{align}
	\begin{split}
		A & = \frac{1}{32\pi^2} \bigg[
			2\lambda_{SR}^2 \, \big(\log\,(2\lambda_{SR}\big)-\tfrac{3}{2})
			-3\yM^4 \,\big(\log\, \yM^2 -\tfrac{3}{2}\big) \bigg] \fineq{,} \\ 
		B & = \frac{2\lambda_{SR}^2 - 3\yM^4}{32\pi^2} \fineq{,}
	\end{split}
	\label{eq:nu_option:A_and_B}
\end{align}
are used. The one-loop thermal contribution $V_\tinytext{FT}$ is given by
\begin{align}
	V_\tinytext{FT}(S,T) & = \frac{T^4}{2\pi^2} \left[
	J_\tinytext{B}\!\left( \frac{m^2_R(S)}{T^2} \right)
    - 6 J_\tinytext{F}\!\left( \frac{m^2_N(S)}{T^2} \right)\right]\,,
		\label{eq:PT:VFT}
\end{align}
with the thermal functions
\begin{align}
J_\tinytext{B,F}(r^2) & = \int_0^\infty \!\! \dd x \; x^2\log\left(1\mp\E^{-\sqrt{x^2+r^2}}\right) \fineq{.}
\end{align}		
Finally, including resummed daisy graphs%
\footnote{As is common in comparable works on phase transitions in theories beyond the Standard Model, we supplement the perturbative expansion of the effective potential with daisy resummation in order to improve the robustness of our results (see \eg \cite{Carrington:1991hz,Kapusta2011}). Note, however, that additional contributions beyond daisy graphs may, in principle, become relevant. Considering such effects is beyond the scope of the present work, though.}
leads to the last term in \cref{eq:PT:Veff}
\begin{align}
V_\text{daisy}(S,T) & = -\frac{T}{12\pi} \left[
\left( m_R^2(S) + \Pi_R(T) \right)^{\nicefrac{3}{2}}
- \left( m_R^2(S) \right)^{\nicefrac{3}{2}}\right]\,,
		\label{eq:PT:Vdaisy}
	\end{align}
with the thermal mass $\Pi_R(T) = (6\lambda_R + \lambda_{SR}) T^2/6$.
Note that we consistently ignored field-independent terms in writing down all of the above expressions.
In all calculations, we will furthermore normalize the effective potential such that it vanishes at \mbox{$S=0$} for all temperatures.

%-----------------------------------------------------------------------------
\section{Phase transition in the model}
\label{sec:PT}
%=======
\noindent
%=======
As we have argued in the last section, one of the distinctive features of the considered model is that the mass scale of right-handed neutrinos must be dynamically generated.
In the following, we will demonstrate, that this necessarily involves a transition from a high-temperature phase with a classically scale-invariant ground state, \mbox{$\chevron{S}=0$}, to a low-temperature phase where scale symmetry is spontaneously broken by quantum fluctuations, \mbox{$\chevron{S}\neq0$}.
Importantly, not only the existence of a scale-symmetry-breaking phase transition (PT), but also its properties are of major interest for gravitational wave phenomenology and will therefore be discussed in the present section.

\begin{figure}[t]
	\centering
	% v(T)
	\subfloat[\label{fig:PT:vT}]{\includegraphics[scale=0.85]{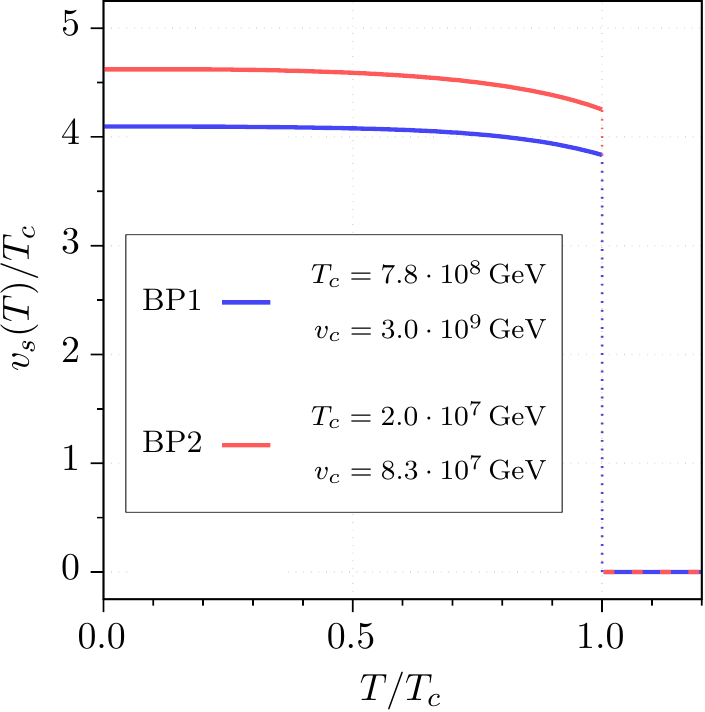}}
	\hspace{4.5em}
	% v_c/T_c in the yM-lSR plane
	\subfloat[\label{fig:PT:vcTc}]{\includegraphics[scale=0.85]{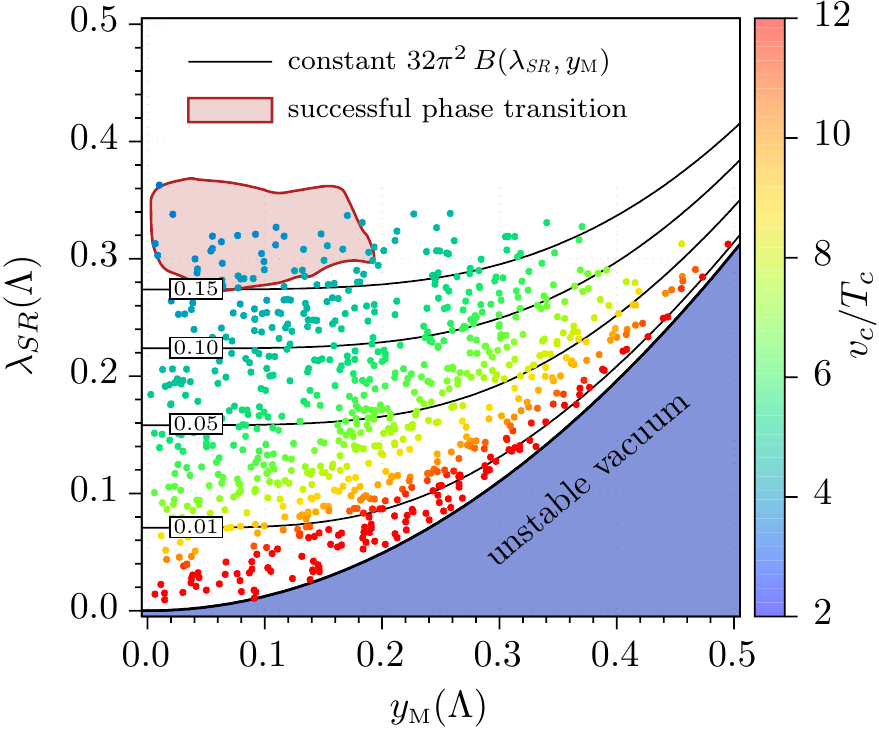}}
	% general
	\caption{Strength of the first-order scale phase transition in our model. (a) Temperature-dependent global minimum of the effective potential for the benchmark points in \cref{tab:PT:benchmark}. (b) Transition strength $v_c/T_c$ (color code) for the comprehensive, exemplary parameter scan presented in Ref.~\cite{Brdar:2018vjq} (\cf Table I and Figure 1 therein for details). All displayed points are fully consistent from the perspective of a zero-temperature analysis as discussed in \cref{sec:nu_option} and Ref.~\cite{Brdar:2018vjq}. However, only for parameter points within the red-shaded area the phase transition was found to actually complete. For further information on how this region was determined, we refer the reader to footnote \ref{fn:PT:vcTc} on page \pageref{fn:PT:vcTc}. Note that the blue-shaded region is excluded since the effective potential becomes unstable at $\Lambda$.}
	\label{fig:PT:groundstate}
\end{figure}

The relevant quantity to explore the model's phase structure more closely is the finite-temperature effective potential $V_\text{eff}$ introduced in \cref{sec:nu_option}.
Based on \cref{eq:PT:Veff}, we can therefore examine how thermal fluctuations change the theory's true groundstate, which is determined by the \textit{global} minimum of $V_\text{eff}$.
To that end, \cref{fig:PT:vT} shows the aforementioned minimum as a function of temperature for the two benchmark points in \cref{tab:PT:benchmark}.
The observed behavior is characteristic for a \textit{first-order} PT, where at a critical temperature $T_c$ there exist two degenerate minima separated by a potential barrier.
Above this temperature, classical scale-invariance is restored by thermal fluctuations, \ie \mbox{$v_s=0$}, whereas the symmetry-breaking minimum \mbox{$v_s\neq0$} is energetically favorable below $T_c$.
% strength of PT for BP
Importantly, \cref{fig:PT:vT} not only reveals the PT's order, but also sheds light on its strength by providing the quantity $v_c/T_c$, where $v_c:=v_s(T=T_c)$ is the non-trivial minimum at the critical temperature.
A first-order PT is typically referred to as \textit{strong} if $v_c/T_c\gtrsim1$, which is satisfied by both of our benchmark points.
% order and strength of PT for all points
In fact, we found that all of the parameter points we investigated a priori suggest the existence of a strong first-order PT, see \cref{fig:PT:vcTc}.
% reference to generic studies
Note that this behavior is expected on general grounds in a model based on classical scale invariance \cite{Konstandin:2011dr}%
\footnote{Even though the given reference focuses on the electroweak phase transition, the arguments remain valid for the model under consideration here.}.

\begin{table}[b]
	\centering
	\sisetup{table-format=1.1e+1,round-mode=figures,round-precision=2}
	\renewcommand{\arraystretch}{1.4}
	\begin{tabular}{c|SSSS|Sc|SS}
		\toprule
		& {\hspace{0.5em}$\GWscale$ [GeV]} & {\symhspace{1.5em}{$\lambda_{SR}$}} & {\symhspace{1.5em}{$\lambda_R$}} & {\symhspace{2em}{$\yM$}} & {\symhspace{1em}{$v_s$ [GeV]}} & \symhspace{0.5em}{$32\pi^2 B$} & {\symhspace{0.7em}{$T_n$ [GeV]}} & {\symhspace{0.7em}{$T_*$ [GeV]}} \\
		\colrule
		\symhspace{0.2em}{BP1} & 1.54852e+09 & 3.20941e-01 & 2.11343e-02 & 9.20936e-02 & 3.17944e+09 & \num{0.20579} & 4.123705e+04 & 1.75469e+08 \\
		\symhspace{0.2em}{BP2} & 4.25894e+07 & 2.95614e-01 & 8.42645e-03 & 1.41853e-01 & 9.02472e+07 & \num{0.173561} & 2.35745e+03 & 4.77299e+06 \\
		\botrule
	\end{tabular}
	\caption{Benchmark points used throughout the paper. All dimensionless couplings are to be understood as $\MSbar$ parameters evaluated at the Gildener-Weinberg scale $\GWscale$. For the corresponding critical temperatures $T_c$, see \cref{fig:PT:vT}. The nucleation and reheating temperatures $T_n$ and $T_*$ are defined in \cref{eq:PT:Tnucl,eq:PT:Trh}, respectively.}
	\label{tab:PT:benchmark}
\end{table}

% bubble nucleation and Hubble expansion
As is well established, a first-order PT proceeds via the nucleation of bubbles of the true vacuum \mbox{$v_s\neq0$}, which then grow inside an expanding universe in the metastable phase, where the system is still at \mbox{$v_s=0$}.
Correspondingly, the transition's actual dynamics is crucially determined by the nucleation rate $\Gamma$ of the aforementioned bubbles, on the one hand, and by the Hubble parameter $H$, on the other hand.
% supercooling
\textit{Strong} first-order phase transitions, as they are predicted by theories based on nearly conformal dynamics, are additionally known to involve a sizable amount of supercooling \cite{Konstandin:2011dr,Megevand2017,Ellis2018}, implying that bubble nucleation can only start at temperatures considerably below the critical one.
Importantly, the energy density in the supercooled phase is then dominated by the energy stored in the scalar field, which entails a (potentially short) epoch of vacuum domination \cite{Konstandin:2011dr}.
In extreme cases, the thus induced inflationary expansion of the universe may even prevent the bubbles from percolating, so that the transition cannot complete \cite{Konstandin:2011dr,Megevand2017,Ellis2018}.
In order to derive robust statements when investigating cosmic phase transitions in a classically scale-invariant model, it is therefore absolutely crucial to carefully take into account the aforementioned vacuum energy contributions.
Still, they are frequently ignored in comparable studies in the literature.

% decay rate of the false vacuum
After these more general remarks, let us now explicitly discuss how to calculate the quantities of interest, starting with the bubble nucleation rate, or, equivalently, the decay rate of the false vacuum $\Gamma$.
Since in our model, the potential barrier separating the true from the false vacuum vanishes at zero temperature, $\Gamma$ can be computed as \cite{Linde:1981zj}
\begin{align}
	\Gamma(T) \simeq T^4 \left( \frac{\Sthree}{2\pi T} \right)^{\nicefrac{3}{2}} \E^{-\Sthree/T} \fineq{,}
	\label{eq:PT:decayRate}
\end{align}
where $\Sthree\equiv \Sthree[S_\text{b}(r)]$ is the three-dimensional Euclidean action of the theory evaluated for the O(3)-symmetric so-called tunneling or bounce solution $S_\text{b}(r)$.
The latter solves the equation of motion for the scalar field $S$,
\begin{align}
	\frac{\dd^2 S}{\dd r^2} + \frac{2}{r}\frac{\dd S}{\dd r} = \frac{\dd V_\text{eff}}{\dd S} \fineq{,}
	\label{eq:PT:eom}
\end{align}
subject to the boundary conditions \mbox{$S\to0$} as \mbox{$r\to\infty$} and \mbox{$\dd S/\dd r=0$} at \mbox{$r=0$} with $r$ being the radial coordinate of three-dimensional space.
% reference to CosmoTransitions
Throughout this work, we use the public code \texttt{CosmoTransitions} \cite{Wainwright2012} to solve \cref{eq:PT:eom}, as well as to compute the Euclidean action $\Sthree$.
% bubble radius check
In doing so, we always check and confirm that, for a given temperature $T$, the corresponding bubble radius%
\footnote{We tend to find rather thick-walled bubbles, for which the bubble radius is not well-defined. As a measure for $r(T)$ we choose the value of $r$, at which the scalar field has dropped to half of its maximum value.}
$r(T)$ satisfies $r(T)\cdot T\gtrsim 1$ \cite{Linde:1981zj}, thus justifying the applicability of \cref{eq:PT:decayRate}.

% Hubble parameter
Let us continue by calculating the Hubble parameter $H$ for some temperature below the critical one.
It is then given via Friedmann's equation in terms of the radiation and vacuum energy densities $\rho_\text{rad}$ and $\rho_\text{vac}$, namely
\begin{align}
	H^2(T) = \frac{\rho_\text{rad}(T) + \rho_\text{vac}(T)}{3\MPl^2} = \frac{1}{3\MPl^2} \left( \frac{\pi^2}{30} g_* T^4 + \Delta V(T) \right)\,.
	\label{eq:PT:Hubble}
\end{align}
A few comments on \cref{eq:PT:Hubble} are in order.
% M_Pl and gEff
First, $\MPl$ denotes the reduced Planck mass, $\MPl=\SI{2.435e18}{GeV}$, while $g_*=\num{106.75}$ is the effective number of relativistic degrees of freedom in the thermal plasma at temperatures below $T_c$ but above the electroweak scale.

% vacuum contribution
Next, the quantity $\Delta V(T):=V_\text{eff}(0,T)-V_\text{eff}(v_s(T),T)$ is the potential difference between the metastable minimum at \mbox{$S=0$} and the stable one at \mbox{$S=v_s(T)\neq0$}.
Thus, it is a measure for the energy stored in the scalar field.
By definition, $\Delta V(T)$ vanishes at the critical temperature and will be small for temperatures \mbox{$T\lesssim T_c$}.
However, as soon as $T\ll T_c$, we expect $v_s(T) \simeq v_s(T=0)$ in accordance with \cref{fig:PT:vT} and $\Delta V(T)$ can be well approximated by its zero-temperature value,
\begin{align}
	\Delta V := \Delta V(T=0) = B v_s^4/2 \fineq{,}
	\label{eq:PT:deltaV}
\end{align}
where we used \cref{eq:PT:Veff,eq:PT:VCW} to find the last relation.
We will apply this approximation throughout the paper.

% T_vac
For the following, it is furthermore instructive to determine the temperature $T_\text{vac}$, below which the Hubble parameter, and thus the expansion of the universe, is dominated by the vacuum term in \cref{eq:PT:Hubble}, \ie
\begin{align}
	\rho_\text{rad}(T_\text{vac}) \stackrel{!}{=} \rho_\text{vac}(T_\text{vac}) \simeq \Delta V
	\quad\quad\Longrightarrow\quad\quad
	T_\text{vac}^4 = \frac{15}{\pi^2 g_*} \cdot B v_s^4 \fineq{,}
	\label{eq:PT:TvacDef}
\end{align}
where we employed \cref{eq:PT:deltaV}. Numerically, the above equation yields
\begin{align}
	T_\text{vac} = \SI[round-mode=figures,round-precision=2]{5.47963e7}{GeV}
				\cdot \biggl( \frac{\tilde{B}}{0.2} \biggr)^{\!\!\nicefrac{1}{4}}
				\cdot \biggl( \frac{v_s}{\SI{e9}{GeV}} \biggr) \fineq{,}
	\label{eq:PT:TvacAna}
\end{align}
where we defined $\tilde{B} := 32\pi^2 B$, which we will see to attain values between \num{0.1} and \num{0.3} for all viable parameter points.
In particular, \cref{eq:PT:TvacAna} therefore demonstrates that $T_\text{vac}$ is always about an order of magnitude smaller than $v_s$ and thus also than the critical temperature, see also \cref{fig:PT:vT}.

% nucleation temperature
With the nucleation rate $\Gamma$ and the Hubble parameter $H$ given in \cref{eq:PT:decayRate,eq:PT:Hubble}, we are now in the position to determine the nucleation temperature $T_n$, which is defined as the temperature for which on average one bubble per horizon volume is produced, \ie
\begin{align}
	\int^{T_c}_{T_n} \frac{\dd T}{T} \frac{\Gamma(T)}{H(T)^4} \stackrel{!}{=} 1 \fineq{.}
	\label{eq:PT:Tnucl}
\end{align}
% Can the PT proceed? --> necessary condition
Obviously, a finite $T_n$ is a necessary condition for the phase transition to proceed.
Parameter points, for which \cref{eq:PT:Tnucl} has no solution, are hence ruled out.
An explicit calculation in our model indeed reveals that only regions of parameter space with moderate PT strength \textendash\ namely $v_c/T_c<5$ \textendash\ are viable in the above sense (\cf the red shaded area%
\footnote{The displayed area was identified by a refined scan in the relevant region of parameter space assuming $\GWscale<\SI{e10}{GeV}$. The associated points are not shown in \cref{fig:PT:vcTc}. The plotted boundary was constructed as the \textit{alpha shape} of the set of all points that are fully consistent from the perspective of a zero-temperature analysis \textit{and} additionally predict a successful PT. In doing so, the shape parameter \mbox{$\alpha=\num{0.03}$} was used (see \eg \cite{alphahull}). \label{fn:PT:vcTc}}
in \cref{fig:PT:vcTc}).
Just as expected, the nucleation temperature for the corresponding points is throughout much smaller than the critical one, see also \cref{tab:PT:benchmark}.

% lower bound for the nucleation temperature
Note furthermore that the consistent implementation of our model as described in \cref{sec:nu_option} requires that the Higgs potential is radiatively generated before the electroweak phase transition takes place.
We will therefore additionally require $T_n > T_\tinytext{EW} = \mathcal{O}(\SI{100}{GeV})$.
However, it turns out that the aforementioned condition is not very restrictive and is automatically satisfied by almost all points for which \cref{eq:PT:Tnucl} can be solved.

% comment percolation temperature
Let us briefly dwell on the issue of whether or not the scale phase transition can complete.
How to answer this question was recently discussed by the authors of Ref.~\cite{Ellis2018}.
To this end, they mainly concentrate on the so-called percolation temperature $T_p$ instead of on the nucleation temperature $T_n$ of \cref{eq:PT:Tnucl}.
Redoing their full calculation of $T_p$ for our model, we only find quantitative but no qualitative differences with respect to $T_n$.
In particular, both temperatures are always of the same order of magnitude and satisfy $T_p \lesssim T_n$.
The authors of Ref.~\cite{Ellis2018} additionally provide a condition (their Eq.~(2.26)) to determine whether the phase transition can actually finish, given a period of inflationary expansion of the universe.
We also checked this criterion to find that for many of our parameter points it is satisfied already at $T_p$, while for almost all other points it is fulfilled at some temperature lower than $T_p$, but above $T_\tinytext{EW}$.

% beta
Once the nucleation temperature is computed via \cref{eq:PT:Tnucl}, a few other quantities relevant for gravitational wave phenomenology can straightforwardly be determined.
First, a measure for the inverse duration of the phase transition is given by
\begin{align}
	\beta := H(T_n) T_n \cdot \frac{\dd}{\dd T}\left( \frac{\Sthree}{T} \right)_{T=T_n} \fineq{.}
	\label{eq:PT:beta}
\end{align}
Numerically, we obtain values of $\beta/H$ ranging between 3 and 10 for all consistent points.
The scale phase transition in our model is therefore predicted to be of relatively long duration, but still fast compared to the expansion of the universe and the corresponding temperature variations.
Note that the observed relations $T_n\ll T_c$ and $\beta/H\gtrsim \mathcal{O}(10)$ characterizing a strongly supercooled yet finite PT can only be satisfied without parameter tuning in models based on nearly conformal dynamics \cite{Konstandin:2011dr}.
% check of exponential approximation
Besides, following Ref.~\cite{Megevand2017} and defining
\begin{align*}
	{\beta^\prime}^2 := \frac{1}{2} H(T_n)^2 \,T_n^2 \cdot \frac{\dd^2}{\dd T^2}\left( \frac{\Sthree}{T} \right)_{T=T_n} \fineq{,}
\end{align*}
we find that $\beta^\prime \gtrsim \text{few } \beta$ holds for all considered viable parameter points, which justifies the applicability of the approximate formulas used to calculate the GW spectra in \cref{sec:GW}.

% alpha
A further crucial quantity is given by the vacuum energy released during the transition normalized to the energy density of the relativistic thermal plasma, \ie
\begin{align}
	\alpha := \frac{\Delta V(T_n)}{\rho_\text{rad}(T_n)}
	\simeq \frac{\Delta V}{\rho_\text{rad}(T_n)}
	\stackrel{\eqref{eq:PT:TvacDef}}{=} \frac{\rho_\text{rad}(T_\text{vac})}{\rho_\text{rad}(T_n)}
	= \frac{T_\text{vac}^4}{T_n^4} \fineq{.}
	\label{eq:PT:alpha}
\end{align}
For the strong scale phase transition in our model, where $T_n\ll T_\text{vac},T_c$, \cref{eq:PT:alpha} explicitly demonstrates that $\alpha$ must always be much larger than one.

\begin{figure}[t]
	\centering
	\includegraphics[scale=0.9]{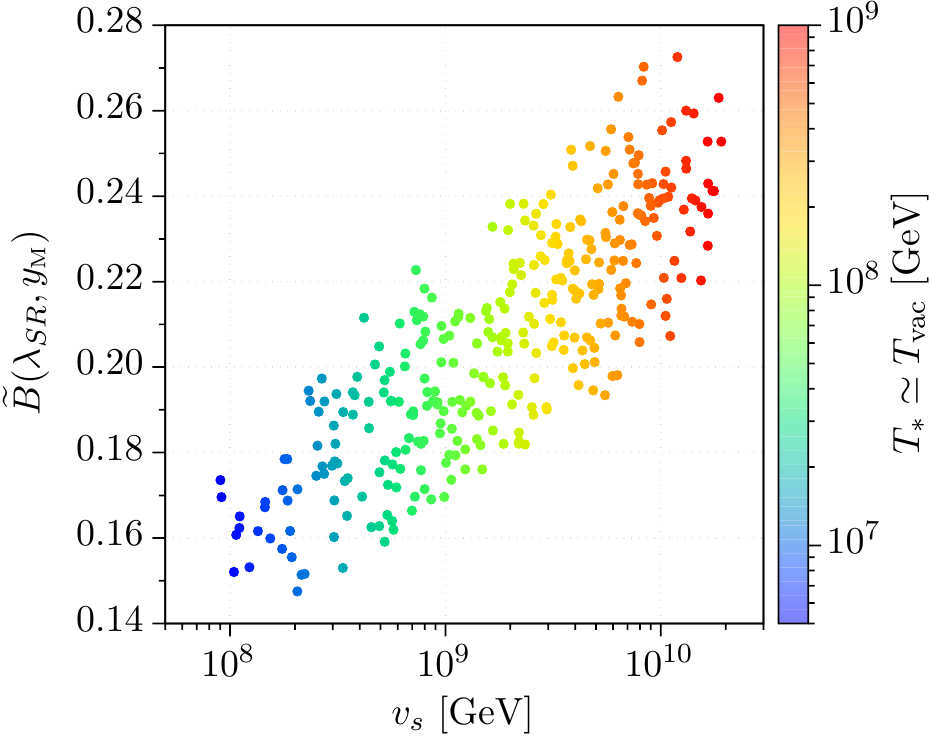}
	\caption{Reheating temperature $T_*$ (color code) in the $v_s$-$\tilde{B}$ plane. The plot explicitly demonstrates that $T_*$ only very mildly depends on $\tilde{B}:=32\pi^2 B$ and that it is well described by \cref{eq:PT:TvacAna}.}
	\label{fig:PT:Trh}
\end{figure}

% reheating temperature
At the end of the phase transition, the aforementioned vacuum energy is initially stored in the gradient of the, now massive, scalon field $S$.
In order to prevent the universe from being matter-dominated for too long after the transition%
\footnote{In particular, the universe has to be radiation-dominated by the time of BBN.},
the model presented in \cref{sec:nu_option} needs to be slightly extended so as to allow the decay of $S$ into Standard Model degrees of freedom.
There exist multiple ways to do so without interfering with any of the model's features described in the previous section.
The simplest scenario is the introduction of a new right-handed neutrino that is lighter than $S$.
In a first step $S$ would then decay to the new fermion species which would subsequently decay into lepton and Higgs doublets.
We have checked that in a substantial portion of this extended model's parameter space the decay rates are larger than the value of the Hubble parameter at $T_n$, implying a fast reheating of the thermal plasma to a temperature $T_*$.
Again, we would like to stress that this is not the only option how to augment the model and achieve a radiation-dominated universe after the PT.
Various minimal extensions of the model could not alone serve for such purpose but also explain the origin of dark matter \cite{Hambye2018a,Baldes2018} and inflation.
Such cases are currently investigated and will be presented in a future publication.

Here, we merely use that for sufficiently fast reheating, energy conservation implies that \mbox{$H(T_n) \simeq H(T_*) \equiv H_*$} and thus \cite{Ellis2018}
\begin{align}
	\rho_\text{rad}(T_*) \simeq \rho_\text{rad}(T_n) 
	+ \rho_\text{vac}(T_n)
	\quad\quad\Longleftrightarrow\quad\quad
	T_* \simeq T_n (1+\alpha)^{\nicefrac{1}{4}} \simeq T_\text{vac} \fineq{,}
	\label{eq:PT:Trh}
\end{align}
where we used \mbox{$\alpha\gg 1$} and \cref{eq:PT:alpha} to obtain the final result.
The reheating temperature as predicted in our model is shown in \cref{fig:PT:Trh}, which shows that it is very nicely described by the analytical formula of \cref{eq:PT:TvacAna}.

%-----------------------------------------------------------------------------
\section{Stochastic gravitational wave signal}
\label{sec:GW}
%=======
\noindent
%=======
After the detailed description of the first-order phase transition in our model, we finally move on to the leading aspect of the present work which is the associated gravitational wave signature.
Before collisions, bubbles are spherically symmetric, so that gravitational waves cannot be produced.
They only arise when bubbles collide via one of three distinct production sources: $(i)$ collisions of shells of the scalar field $S$ \cite{Kosowsky:1991ua}, $(ii)$ sound waves \cite{Hindmarsh:2013xza}, and $(iii)$ magnetohydrodynamic turbulences \cite{Caprini:2006jb} in the plasma following bubble collisions.
In our model \mbox{$\alpha\gg 1$} and, as discussed in \cref{sec:PT}, the phase transition happens during the vacuum-dominated epoch \cite{Ellis2018}.
Then, it is well-known that gravitational waves are dominantly produced from collisions of the bubble walls \cite{Caprini:2015zlo} while the other mechanisms can be safely ignored.
In order to estimate the gravitational wave signal's strength we employ results from numerical simulations which use the so called \enquote{envelope approximation} \cite{Huber:2008hg}. The stochastic gravitational wave spectrum is
\begin{align}
	\Omega_\tinytext{GW}(f)\, h^2 = \num{1.67e-5} \left(\frac{\beta}{H_*}\right)^{\!\!-2} \left(\frac{\kappa \, \alpha}{1+\alpha}\right)^{\!2} \left( \frac{100}{g_*} \right)^{\!\!\nicefrac{1}{3}} \left(\frac{0.11\, v_w^3}{0.42+v_w^2}\right) \frac{3.8 \,(f/f_\text{peak})^{2.8}}{1+2.8 \,(f/f_\text{peak})^{3.8}}\,,
	\label{eq:GW-spectrum}
\end{align}
where $\kappa$ is the fraction of the vacuum energy that is converted into the gradient energy of the $S$ field, $v_w$ is the bubble wall velocity and the gravitational wave peak frequency reads \cite{Kamionkowski:1993fg}
\begin{align}
	f_\text{peak}= \num{16.5e-6} \left(\frac{\beta}{H_*}\right) \left(\frac{0.62}{1.8-0.1 \, v_w + v_w^2}\right) \left(\frac{T_*}{100\,\text{GeV}}\right) \left( \frac{g_*}{100} \right)^{\!\!\nicefrac{1}{6}}\, \text{Hz}\,.
	\label{eq:peak}
\end{align}
In our model \mbox{$\kappa\approx 1$} and the bubble wall velocity approaches the speed of light, \ie \mbox{$v_w\to1$} \cite{Bodeker:2009qy}.
Given such high velocities as well as the dominance of the vacuum energy  at the phase transition, the scenario under our consideration is usually referred to as \enquote{runaway bubbles in vacuum} \cite{Espinosa:2010hh}. Let us note that the gravitational wave spectrum  depends only on $\beta$ and $T_*$ since any $\alpha$ dependence drops out in the \mbox{$\alpha\gg 1$} limit.
In \cref{sec:PT} we have discussed how these parameters are obtained and here we will present results for all sampled parameter points which are consistent with a successful phase transitions, as well as with the basic requirements of the model (see \cref{sec:nu_option} or Ref.~\cite{Brdar:2018vjq}).

\begin{figure}[t]
	\centering
	\includegraphics[scale=0.9]{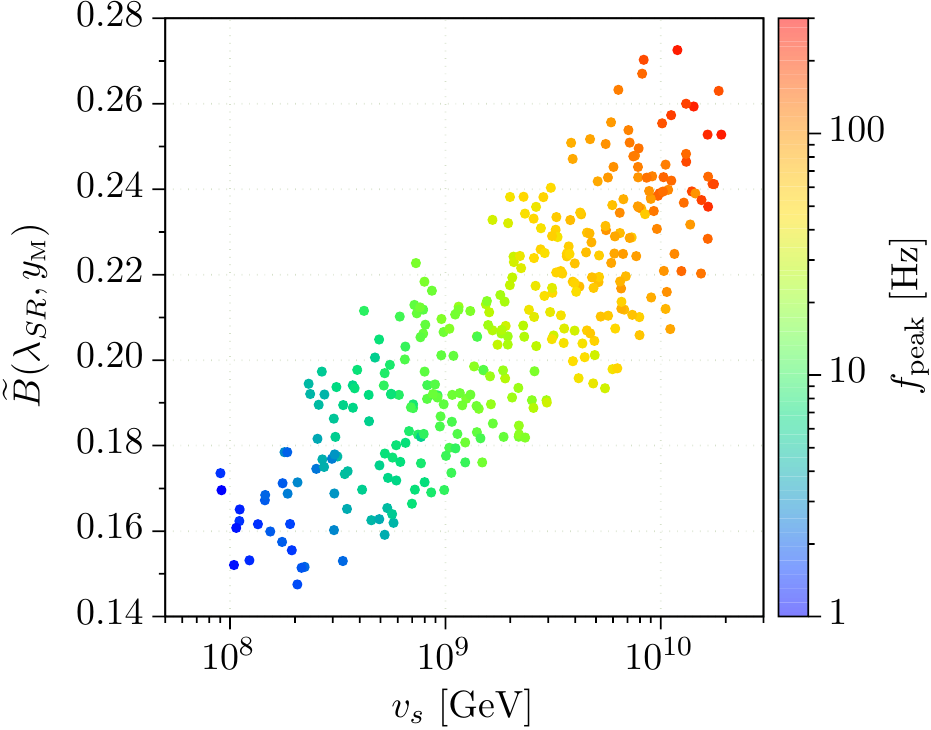}
	\caption{Values of  $f_\text{peak}$ (color code) presented in the $v_s$-$\tilde{B}$ plane. Generally, for larger $T_*$, the peak frequency increases as can be seen from \cref{eq:peak}. The peak frequency in our model ranges from $\mathcal{O}(\SI{1}{Hz})$ to $\mathcal{O}(\SI{100}{Hz})$.}
	\label{fig:fpeak}
\end{figure}

We have shown in \cref{eq:PT:TvacAna} that $T_*$ can be expressed in terms of $\tilde{B}$ and $v_s$, which is also numerically demonstrated in \cref{fig:PT:Trh} for the parameter points that are compatible with the scale phase transition's successful completion.
In \cref{fig:fpeak} we show the peak frequency $f_\text{peak}$ in the $v_s$-$\tilde{B}$ parameter space.
This figure seemingly does not qualitatively differ from \cref{fig:PT:Trh} where $T_*$ is shown in the same plane.
However, while in \cref{fig:PT:Trh} one can see a clear separation between points of different color (\ie different $T_*$), which simply stems from the aforementioned analytical relation in \cref{eq:PT:TvacAna}, this is not the case in \cref{fig:fpeak}.
There, overlapping points of different color arise due to the additional parameter appearing in \cref{eq:peak}, namely the inverse of the phase transition's duration $\beta$, which introduces a mild but non-trivial dependence on the model parameters $\yM$, $\lambda_{SR}$ and $\lambda_{R}$.
For our viable parameter points, the values of $\beta/H_*$ indicate a reasonably fast phase transition.

%%%%%%%%%%%%%%%%%%%%%%%%%%%%%%%%%%%%%%%%%%%%%%%%%%%%%%%%
%%%%%%%%%%%%%%%%%%%%%%%%%%%%%%%%%%%%%%%%%%%%%%%%%%%%%%%%

After reviewing the spectrum of stochastic gravitational waves in our model (see \cref{eq:GW-spectrum}), we would now like to quantify the discovery potential at relevant current and near-future gravitational wave detectors for each viable parameter point in our model.
As representatives for \textit{ground-based} observatories we study LIGO \cite{Abbott2016} and Virgo \cite{TheVirgo:2014hva}.
The observing runs which we consider are O2 (2016--17), O3 (2019--20) and \enquote{Design} (2022+), where in brackets we denote the years in which a given run is conducted.
As we have explicitly checked, the overall sensitivity of the LIGO-Virgo network to stochastic background in all of the aforementioned phases is dominated by the two LIGO detectors.
In what follows, we will therefore mostly refer to LIGO, although Virgo contributions are included in our calculations.
Currently, LIGO is upgrading the detectors and will commence its O3 stage in February 2019.

Apart from LIGO and Virgo, we also confront the gravitational wave spectra from our model with the sensitivities of \textit{space-based} detectors, namely LISA \cite{Caprini:2015zlo}, Big Bang Observer (BBO) \cite{Corbin:2005ny}, and DECIGO (two stages: B-DECIGO and FP-DECIGO \cite{Seto:2001qf}). Following the literature, we define the signal-to-noise ratio (SNR) as (see \eg \cite{Thrane:2013oya})
\begin{align}
	\text{SNR}=\sqrt{2 t_{\text{obs}} \int_{f_{\text{min}}}^{f_{\text{max}}} \!\! \dd f \, \bigg[
\frac{\Omega_\tinytext{GW}(f)\, h^2}{\Omega_\text{noise}(f)\, h^2}\bigg]^2},
	\label{eq:SNR}
\end{align}
where $t_{\text{obs}}$ denotes the duration of an observation in seconds, while $f_\text{min}$ and $f_\text{max}$ define the experiment's bandwidth.
The quantity $\Omega_\text{noise}(f)\,h^2$ represents the effective strain noise power spectral density for a given detector network, expressed as energy density parameter \cite{Moore:2014lga}.
For the space-based observatories mentioned above, we adopt the strain noise power spectral densities from Refs.~\cite{Yagi:2011yu,Yagi:2013du,Isoyama:2018rjb}.
For LIGO and Virgo, on the other hand, we follow Ref.~\cite{Aasi:2013wya}.
Calculating $\Omega_\text{noise}$ generally also requires knowledge of the frequency-dependent normalized isotropic overlap reduction functions $\gamma$ for each detector pair in the network.
For the LIGO-Virgo network the $\gamma$ functions can be found \eg in Ref.~\cite{Romano2017}.
For all space-based experiments we assume \mbox{$t_{\text{obs}}=\SI{5}{years}$}, while the durations of the different LIGO runs are set according to Refs.~\cite{Abbott:2017xzg,Aasi:2013wya} (for each LIGO stage we additionally assume a duty cycle of \SI{50}{\%} \cite{Abbott:2017xzg}).
Note that the expression for SNR in \cref{eq:SNR} must be multiplied by a factor of $1/\sqrt{2}$ for LISA and B-DECIGO to account for the fact that these experiments perform autocorrelation measurements based on a \textit{single} detector in order to search for stochastic gravitational waves.
Let us furthermore remark that, in principle, the contribution of unresolvable astrophysical foregrounds (neutron star, black hole, white dwarf mergers) should be included in the denominator of \cref{eq:SNR}. However, it was shown in Ref.~\cite{Rosado:2011kv} that there is no unresolvable foreground in the \mbox{$f>\mathcal{O}(\SI{1}{Hz})$} frequency range which is preferred by the parameter points in our model.

%%%%%%%%%%%%%%%%%%%%%%%%%%%%%%%%%%%%%%%%%%%%%%%%%%%%
\begin{figure}[t]
	\centering
	\includegraphics[scale=0.4]{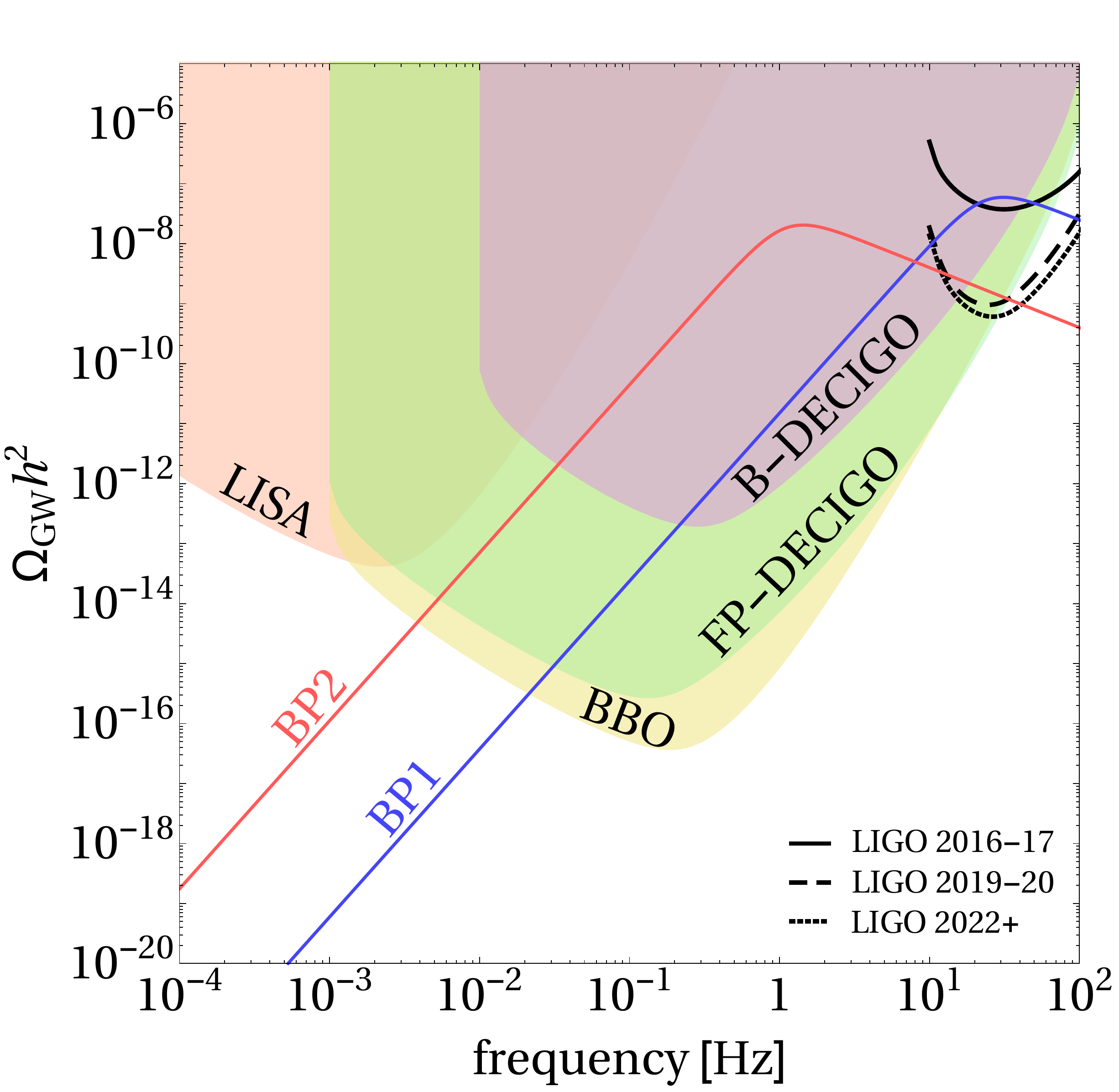}
	\caption{Stochastic gravitational wave spectra for the benchmark points BP1 and BP2 (see \cref{tab:PT:benchmark}) are shown together with the limits (LIGO 2016--2017) and future sensitivities (LIGO 2019--2020, LIGO 2022+, DECIGO, BBO, LISA) of selected observatories. Our model predicts values for $f_\text{peak}$ in the $\mathcal{O}(\SI{1}{Hz})$ to $\mathcal{O}(\SI{100}{Hz})$ range (see \cref{fig:fpeak}) with an associated peak energy density of order \num{e-8}. Hence, the majority of the parameter points can already be tested in the upcoming LIGO observing run. The displayed power-law integrated sensitivity curves were constructed according to Ref.~\cite{Thrane:2013oya} assuming \mbox{$\text{SNR}_\text{thr}=10$} for space-based experiments, and \mbox{$\text{SNR}_\text{thr}=2$} for LIGO, respectively. Further information can be found in the main text below \cref{eq:SNR}.}
	\label{fig:spectrum}
\end{figure}
%%%%%%%%%%%%%%%%%%%%%%%%%%%%%%%%%%%%%%%%%%%%%%%%%%%%

Sensitivity contours%
\footnote{To be precise, \cref{fig:spectrum} displays power-law integrated (PI) sensitivity curves, which differ from the strain noise power spectral densities entering in \cref{eq:SNR}. The procedure of how to construct PI curves is outlined in \cite{Thrane:2013oya}.}
for all studied experiments are plotted in \cref{fig:spectrum} where we also show the gravitational wave spectra $\Omega_\tinytext{GW}(f)\,h^2$ for the two benchmark points (BP1 and BP2) given in \cref{tab:PT:benchmark}.
Clearly, LIGO is most sensitive in the \SIrange{10}{100}{Hz} frequency range, whereas the space-based detectors will be able to also probe smaller frequencies.
The stochastic gravitational wave signal for the parameter point BP1 marginally intersects the O2 sensitivity contour and the corresponding SNR is \num{2.7}.
Importantly, BP1 nicely demonstrates that parameter points in our model can have $f_\text{peak}$ at frequencies where LIGO is most sensitive.
Furthermore, it can be easily obtained from \cref{eq:GW-spectrum} that for a typical value \mbox{$\beta/H_*\sim 10$}, $\Omega_\tinytext{GW} (f_\text{peak})\,h^2$ is around \num{e-8}.
This can also be seen in \cref{fig:spectrum} for both benchmark points. Therefore, by looking at the sensitivity curves from \cref{fig:spectrum}, one may infer that all parameter points with $f_\text{peak}$ in the \SIrange{10}{100}{Hz} region will be tested in LIGO's O3 and \enquote{Design} phases.
In contrast, due to a peak frequency of only roughly \SI{1}{Hz}, BP2 is much less likely to be successfully probed by LIGO.
However, both considered stages of the future space-based observatory DECIGO, as well as the BBO experiment would still be sensitive.

The shape of the gravitational wave spectrum can be easily understood from \cref{eq:GW-spectrum}.
Namely, for \mbox{$f\ll f_\text{peak}$}, $\Omega_\tinytext{GW}(f)\,h^2$ is proportional to $f^{2.8}$, whereas in the high-frequency region, \mbox{$f\gg f_\text{peak}$}, the gravitational wave signal drops less steeply $\Omega_\tinytext{GW}(f)\,h^2 \propto f^{-1}$.
This can also be inferred from \cref{fig:spectrum} for both of our benchmark points.
Let us note that even though causality implies that the signal should increase with the third power in the low-frequency limit \cite{Caprini:2009fx}, $f^{2.8}$ yields a better fit to the simulated data and is therefore commonly employed in the literature \cite{Caprini:2015zlo}.

%%%%%%%%%%%%%%%%%%%%%%%%%%%%%%%%%%%%%%%%%%%%%%%%%%%%
\begin{figure}[t]
	\centering
	\captionsetup[subfigure]{oneside,margin={3.5em,0em}}
	% SNR > 10
	\subfloat[\label{fig:ligoA}]{\includegraphics[scale=0.85]{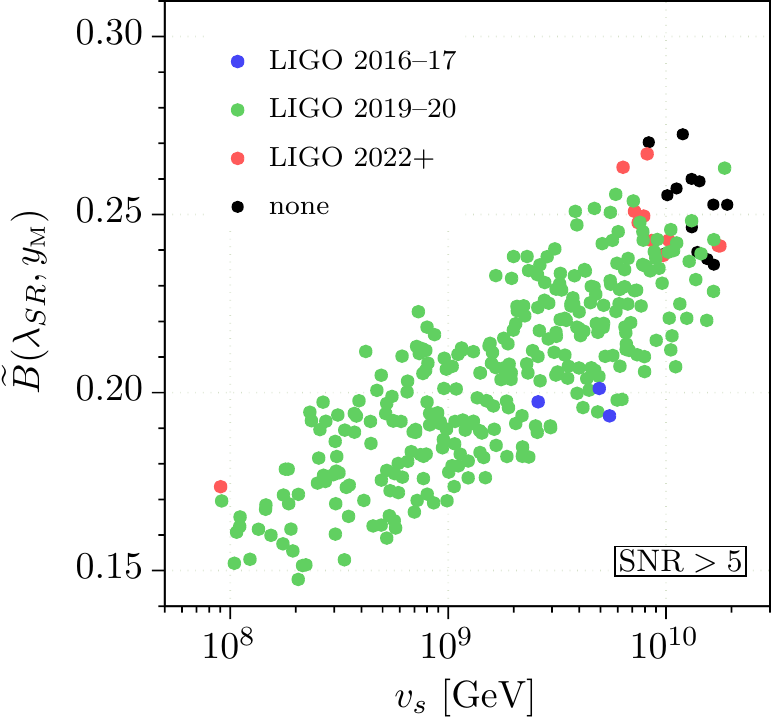}}
	\hspace{4.5em}
	% SNR > 30
	\subfloat[\label{fig:ligoB}]{\includegraphics[scale=0.85]{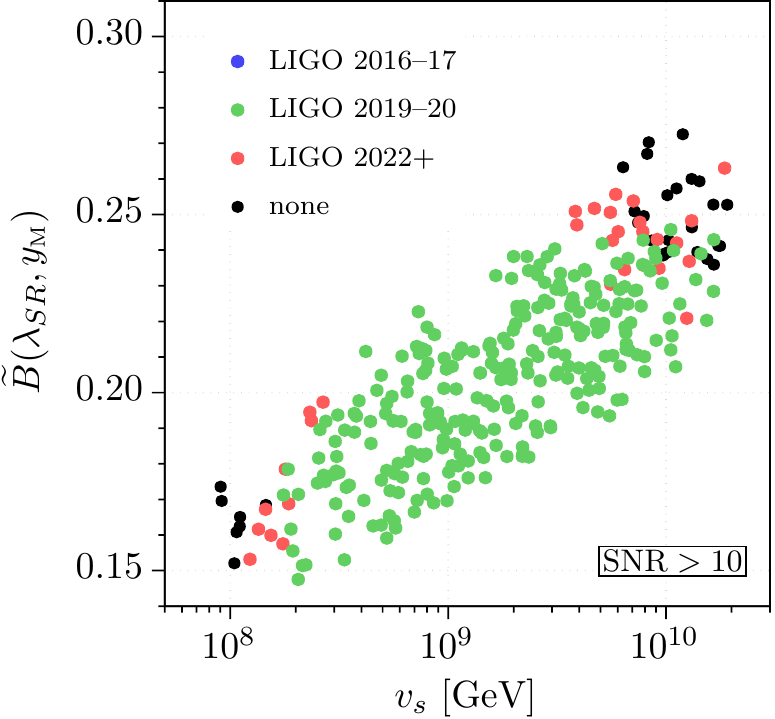}}
	% general
	\caption{In the figure, all the viable points in our parameter scan are shown. For a given phase of LIGO (color code) we indicate in the left (right) panel 
	the points for which the signal-to-noise ratio SNR as defined in \cref{eq:SNR} exceeds $5$ $(10)$. If more than one phase satisfy the requirement, the color corresponds to the earliest one.}
	\label{fig:LIGO}
\end{figure}
%%%%%%%%%%%%%%%%%%%%%%%%%%%%%%%%%%%%%%%%%%%%%%%%%%%

Determining the precise value of SNR for which a stochastic gravitational wave background can be reliably detected in a given experiment (hereafter denoted as $\text{SNR}_\text{thr}$) is beyond the scope of the present paper as it is related to experimental aspects which we do not study.
Following Ref.~\cite{Caprini:2015zlo} as a representative work on space-based detectors (see also recent \cite{Aoki:2017aws,Madge:2018gfl} and references therein), as well as Ref.~\cite{Abbott:2017xzg} for LIGO, we infer that \mbox{$\text{SNR}_\text{thr}=10$} may be generally regarded as a conservative estimate.
For experiments that have several phases such as LIGO and DECIGO we simply calculate SNR values for each phase independently, without taking into account cumulative effects.
This already robustly demonstrates during which phase a given parameter point can be tested.
In particular, since LIGO's O2 phase is much less sensitive than O3, the combination of the two does not yield significant improvement with respect to O3-only.
The left (right) panel of \cref{fig:LIGO} shows for all viable parameter points and for the aforementioned observing runs whether SNR exceeds 5 (10).
The color code is indicated in the figures.
If multiple phases of the experiment satisfy the given SNR requirement, the parameter point is plotted in the color associated with the earliest phase.
For instance, the blue points in the left  panel of \cref{fig:LIGO} denote parameter points which produce $\text{SNR}>5$ in LIGO's O2 run.
Clearly, SNR in the O3 and \enquote{Design} stages will then automatically exceed 5, yet the color corresponds to the O2 phase.
The appearance of black points in both panels signals that the considered SNR cannot be reached by any phase individually.
If \mbox{$\text{SNR}_\text{thr}=10$} is indeed a realistic signal-to-noise ratio for an actual discovery of stochastic gravitational waves, we infer from \cref{fig:ligoB} that a large portion of parameter space (about $\SI{85}{\%}$ of all viable points) will be tested in LIGO's forthcoming O3 phase.
Let us note that our conclusion is effectively unaltered if we set \mbox{$\text{SNR}_\text{thr}=5$}, as can be seen from \cref{fig:ligoA}.
Lastly, there are only few parameter points (blue color) that can already be excluded by the O2 run assuming \mbox{$\text{SNR}_\text{thr}=5$} is appropriate.

By comparing \cref{fig:fpeak,fig:LIGO}, we learn that the parameter points which will be hard to test in O3 exhibit peak frequencies that are either \mbox{$f_\text{peak}<\SI{10}{Hz}$} or  \mbox{$f_\text{peak}\gtrsim \SI{100}{Hz}$}.
In these regions, LIGO's sensitivity is weaker as can be deduced from the black curves in \cref{fig:spectrum}. This is also seen by comparing the left and right panels of \cref{fig:LIGO}:
Increasing the considered SNR threshold value from $5$ to $10$, several points at  both edges (with highest and lowest peak frequencies)  turn black, \ie their SNR does not exceed 10 for any of the LIGO stages.
 
%%%%%%%%%%%%%%%%%%%%%%%%%%%%%%%%%%%%%%%%%%%%%%%%%%%%%%%%
\begin{figure}[t]
	\centering
	\includegraphics[scale=0.9]{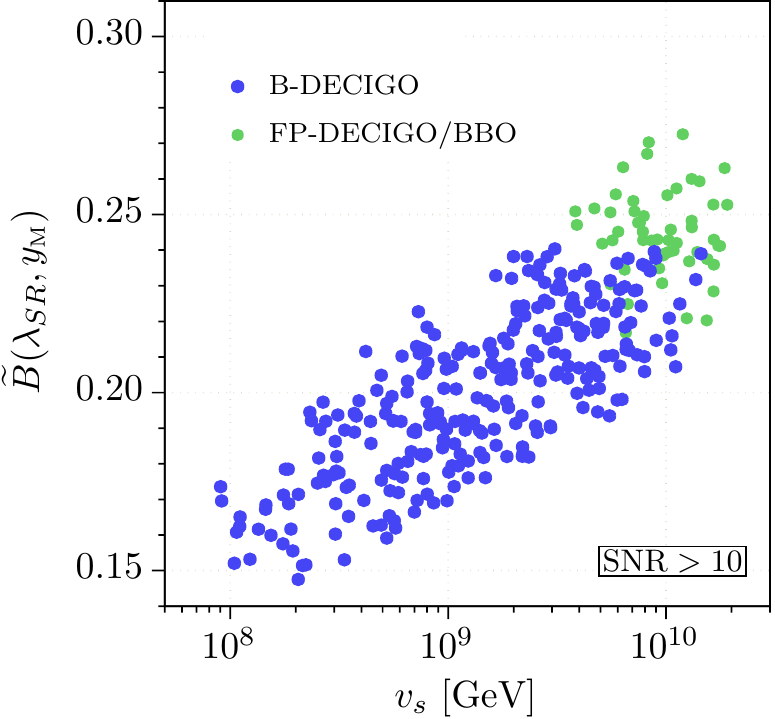}
	\caption{Same as \cref{fig:LIGO}, but for space-based experiments. Blue points are predicted to produce \mbox{$\text{SNR}>10$} in a five-year B-DECIGO run. The remaining points (green) are expected to yield \mbox{$\text{SNR}>10$} in both FP-DECIGO and BBO runs of the same length. In fact, the two aforementioned observatories can fully test the considered model even with \mbox{$t_\text{obs}<\SI{1}{year}$}.}
	\label{fig:DECIGO}
\end{figure}
%%%%%%%%%%%%%%%%%%%%%%%%%%%%%%%%%%%%%%%%%%%%%%%%%%%%%%%% 

In analogy to \cref{fig:LIGO}, the plot in \cref{fig:DECIGO} shows the \mbox{$\text{SNR}_\text{thr}=10$} case for the considered DECIGO stages (early B-DECIGO and FP-DECIGO).
Again, using the sensitivities in \cref{fig:spectrum} it is straightforward to understand this figure. The FP-DECIGO phase is more sensitive in all frequency regions.
In particular, for peak frequencies larger than about \SI{100}{Hz} we observe green points indicating testability only by FP-DECIGO.
This is because for highest peak frequencies the gravitational wave spectrum only marginally intersects B-DECIGO's sensitivity curve, while it significantly overlaps with FP-DECIGO's.
Hence, B-DECIGO will probe points with lower peak frequencies, whereas FP-DECIGO is sensitive to the full parameter space assuming \mbox{$\text{SNR}_\text{thr}=10$} as a realistic detection threshold.
The same is true for BBO whose sensitivity at least matches that of FP-DECIGO for all frequencies, while around \mbox{$\mathcal{O}(\SI{1}{Hz})$} it even yields a significant improvement.
In contrast, LISA is not particularly sensitive (\mbox{$\text{SNR}_\text{thr}>10$} is not achieved for any parameter point) because frequency-wise LISA is more appropriate to test TeV-scale new physics.

In conclusion, after reviewing the stochastic gravitational wave spectra and methods to quantify their testability, we demonstrate that our model can be probed by both ground- and space-based observatories.
Given the preference for high scales (\mbox{$v_s>\num{e7}\text{GeV}$}), detection of gravitational waves from the first-order scale phase transition is essentially the only option to test the considered model since we cannot access such high energies in any conventional particle physics experiment.

%-----------------------------------------------------------------------------
\section{Summary and Conclusions}
\label{sec:conclusion}
%-----------------------------------------------------------------------------
\noindent
In this paper we further investigated the framework introduced in Ref.~\cite{Brdar:2018vjq}.
There, the authors proposed a UV-complete, scale-invariant model in which both neutrino masses and the Higgs potential originate from the effects of right-handed neutrinos whose Majorana mass terms dynamically develop through the spontaneous breaking of classical conformal symmetry.
The heavy right-handed neutrinos then give rise to nonzero active neutrino masses through the type-I seesaw mechanism, while the Higgs potential is generated at the quantum level, namely from the loops in which the neutrinos propagate.

In the present work, in addition to the Coleman-Weinberg effective potential, we considered  leading thermal contributions for the purpose of a detailed investigation of the phase transition associated with the breaking of the model's scale symmetry. We found that the phase transition is of strong first order,  which implies a significant supercooling. We inferred that the universe becomes vacuum-dominated at temperatures 
 which are roughly an order of magnitude smaller than the critical one. The vacuum energy dominance induces an epoch of exponential expansion of the universe which renders bubble nucleation more difficult. In particular, we found that there is a  large portion of the model's parameter space which can be excluded due to the failure of the scale phase transition.

For the remaining parameter points which yield a successful phase transition, we have considered a phenomenological signature: the stochastic gravitational wave background induced by bubble collisions. Since the lower bound on the VEV of the scalar field that breaks scale invariance is \SI{e7}{GeV}, the typical gravitational wave frequencies are much larger in comparison to those associated with TeV-scale phase transitions. For the latter, space-based detectors such as LISA are most sensitive, whereas
our model can be robustly probed by ground-based detectors.
Remarkably, the upcoming science run of LIGO will test practically the full parameter space of the considered model. If this model was chosen by Nature, then the gravitational wave signature is the only method which can lead to a successful discovery as the new physics particles are too massive to be produced in current and near-future colliders.

\begin{acknowledgments}
\noindent
We would like to thank Manfred Lindner, Thomas Konstandin, Ville Vaskonen and Susan van der Woude for useful discussions, as well as Carroll L.~Wainwright for correspondence regarding the \texttt{CosmoTransitions} package.
Furthermore, we are deeply indebted to Andrew Long, Maria Alessandra Papa and Andrew Matas for most helpful assistance in obtaining the correct LIGO sensitivities, as well as to Takashi Nakamura and Hiroyuki Nakano for providing information on the DECIGO sensitivities.
The work of JK is partially supported by the Grant-in-Aid for Scientific Research (C) from the Japan Society for Promotion of Science (Grant No.16K05315).
\end{acknowledgments}

\bibliographystyle{JHEP}
\bibliography{refs}
%-----------------------------------------------------------------------------

\end{document}